\documentclass[aps,prl,reprint,groupedaddress]{revtex4-1}
\usepackage{graphicx} 
\usepackage{dcolumn} 
\usepackage{bm}
\usepackage{hyperref}

\begin{document}

\title{ Effect of cluster transfer on neutron-rich nuclide production around N=126 in multinucleon transfer reactions }
\author{Zhao-Qing Feng}
\email{Corresponding author: fengzhq@scut.edu.cn}

\affiliation{School of Physics and Optoelectronics, South China University of Technology, Guangzhou 510640, China }

\date{\today}

\begin{abstract}

The cluster transfer in multinucleon transfer reactions near Coulomb barrier energies is implemented into the master equations in dinuclear system model, in which the deuteron, triton, $^{3}$He and $\alpha$ are taken into account. The effects of cluster transfer and dynamical deformation on the formation of primary and secondary fragments are systematically investigated. It is found that the inclusion of cluster transfer is favorable the fragment formation with increasing the transferring nucleons and leads to a broad mass distribution. The isotopic cross sections of elements W, Os, Rn and Fr in the reaction of $^{136}$Xe+$^{208}$Pb at the incident energy of E$_{c.m.}$ = 450 MeV are nicely consistent with the Argonne data. The new neutron-rich isotopes of wolfram and osmium are predicted with cross sections above 10 nb. The production mechanism of neutron-rich heavy nuclei around N = 126 in the reactions of $^{58,64,72}$Ni + $^{198}$Pt is investigated thoroughly. The cross sections for producing the neutron-rich isotopes of platinum, iridium, osmium and rhenium in the multinucleon transfer reactions of $^{64}$Ni + $^{198}$Pt and $^{72}$Ni + $^{198}$Pt at the center of mass energies of 220 MeV and 230 MeV are estimated and proposed for the future experiments.
\begin{description}
\item[PACS number(s)]
25.70.Hi, 25.70.Lm, 24.60.-k      
\end{description}
\end{abstract}

\maketitle

The completeness of element periodic table and heavy element origin in the universe are the basic science problems \cite{Sc05}, which are associated with the synthesis of superheavy element, rapid-neutron capture process in the big-bang nucleosynthesis, shell evolution, nuclear fission etc. In the terrestrial laboratories, the neutron-rich heavy or superheavy nuclei might be created via different ways, i.e., the projectile fragmentation reactions, spallation and fission reactions of heavy nuclei, fusion-evaporation reaction, multinucleon transfer reaction etc. The fusion-evaporation reactions, namely, the cold fusion reactions with $^{208}$Pb or $^{209}$Bi based targets \cite{Og75,Ho00,Mo04,Mu15} and the $^{48}$Ca induced warm reactions \cite{Og99,Og06,Og15}, have been extensively used for synthesizing the superheavy nucleus (SHN). The heavy nuclides created by the fusion-evaporation reactions are located in the neutron-deficient regime in the nuclear chart and away from the 'stability of island' \cite{My66,So66}. Up to now, roughly 3200 nuclides are created in different laboratories in the world via the projectile fragmentation, spallation and fission reactions of heavy nuclei, fusion-evaporation reaction, transfer reaction etc \cite{Th16}. Recently, the multinucleon transfer (MNT) reaction attracted much attention for producing the neutron-rich heavy and superheavy nuclei. With constructing the new facilities in the world such as RIBF (RIKEN, Japan) \cite{Sa18}, SPIRAL2 (GANIL in Caen, France) \cite{Ga10}, FRIB (MSU, USA) \cite{We19}, HIAF (IMP, China) \cite{Ch19}, the SHNs on the 'island of stability' by using the neutron-rich radioactive beams induced fusion reactions or via the multinucleon transfer (MNT) reactions might be created in experiments. The spectroscopic measurements of neutron-rich heavy nuclei around N=126, 152 and 162 are particular important for understanding the single-particle motion, shape coexistence, new decay mode beyond the binary fission etc.

Since 1970s, the MNT reactions or deep inelastic heavy-ion collisions were extensively investigated in experiments, in which the new neutron-rich isotopes of light nuclei and also proton-rich actinide nuclei were observed \cite{Art71,Art73,Art74,Hil77,Gla79,Moo86, Wel87}. The nuclear dynamics in the MNT reactions were thoroughly investigated, i.e., the fragment cross section, total kinetic energy spectra of fragments, angular distribution, relative motion energy and angular momentum dissipation, two-body kinematics, etc. Recently, the MNT reactions attracted attention again in experiments for the new isotope production. It has been manifested that the MNT reactions have the advantage of broad isotope distribution, e.g., more than 100 nuclides with \emph{Z}=82-100 in the reaction of $^{48}$Ca + $^{248}$Cm and five new neutron-deficient isotopes $^{216}$U, $^{219}$Np, $^{223}$Am, $^{229}$Am and $^{233}$Bk \cite{De15}. Recently, the MNT reaction mechanism was investigated both in experiments and in theories around the neutron shell closure of N = 126. The production cross sections, total kinetic energy spectra and angular distributions were measured in the reactions of $^{136}$Xe + $^{208}$Pb \cite{Ko12,Ba15}, $^{136}$Xe + $^{198}$Pt \cite{YX15,De19}, $^{156,160}$Gd + $^{186}$W \cite{EM17}, and $^{238}$U + $^{232}$Th \cite{SK18}. Recently, the new isotope $^{241}$U was created in the MNT reactions of $^{238}$U + $^{198}$Pt \cite{Ni23}. The neutron-rich nuclides around the neutron shell closure N = 126 have significant application in understanding the origin of heavy elements from iron to uranium in the \emph{r}-process of nucleosynthesis in the stellar evolution. It has been confirmed that the shell closure plays an important role on the production of neutron-rich nuclei and more advantage with the multinucleon transfer (MNT) reactions in comparison to the projectile fragmentation \cite{Ku14}. The MNT reactions with the neutron-rich radioactive beams have more advantage for creating the rare isotopes beyond the $\beta$-stability line.

Several models have been proposed for describing the MNT reactions, i.e., the dinuclear system (DNS) model \cite{Fe09,GG10,GG81}, the GRAZING model \cite{Win94,Win95}, the dynamical model based on multidimensional Langevin equations \cite{Zag15,Ka17,Sa19} etc. Moreover, the microscopic approaches based on the nucleon degree of freedom, the time-dependent Hartree-Fock (TDHF) approach \cite{CC09,KK13,WN18,Gu19} and extension by incorporating fluctuation and correlation in the nuclear transfer based on the stochastic mean-field theory \cite{Se20}, the improved quantum molecular dynamics (ImQMD) \cite{TJ08,ZK13,ZK15} are also used to describe the MNT reactions. Some interesting issues have been investigated with the models, e.g., the production cross sections of new isotopes, total kinetic energy spectra and polar angle distribution of transfer fragments, structure effect on the fragment formation etc. There are still some open problems and model improvements are necessary for the MNT reactions, i.e., the preequilibrium cluster emission, the stiffness of nuclear surface during the nucleon transfer process, the massive transfer dynamics, coupling of the radial motion and neck evolution, etc.

In the letter, the cluster transfer in solving the master equations has been implemented into the DNS model, in particular for transferring the deuteron, triton, $^{3}$He and $\alpha$. The MNT dynamics is to be investigated with the DNS-cluster model. In the DNS model, the nucleon transfer between the binary fragments is governed by the single-particle Hamiltonian \cite{Fe07}. Only the nucleons within the valence space are active for transfer \cite{W24,S25}. The transition probability is related to the local excitation energy and nucleon transfer \cite{Fe07,Fe06}, which is microscopically derived from the interaction potential in valence space. The local excitation energy is determined by the dissipation energy from the relative motion and the potential energy surface of the DNS. The dissipation of the relative motion and angular momentum of the DNS is described by the classical trajectory with damped collisions. The cross sections of the primary fragments (\emph{Z}$_{1}$, \emph{N}$_{1}$) are calculated as follows:
\begin{eqnarray}
	\sigma_{pr}(Z_{1},N_{1},E_{c.m.}) = && \sum^{J_{max}}_{J=0} \sigma_{cap}(E_{c.m.},J) \int f(B)     \nonumber\\
	&&  \times P(Z_{1},N_{1},E_{1},J_{1},B)dB
\end{eqnarray}
The secondary decay of the primary fragments is considered to form the final MNT fragments. The cross section is evaluated by
\begin{eqnarray}
	\sigma_{sur}(Z_{1},N_{1},E_{c.m.}) = && \sum^{J_{max}}_{J=0} \sigma_{cap}(E_{c.m.},J) \int f(B)   \nonumber\\
	&&  \times \sum_{s} P(Z^{\prime}_{1},N^{\prime}_{1},E^{\prime}_{1},J^{\prime}_{1},B)                   \nonumber \\
	&&  \times W_{sur}(Z^{\prime}_{1},N^{\prime}_{1},E^{\prime}_{1},J^{\prime}_{1},s)dB.
\end{eqnarray}
Here, $E_{1}$ and $J_{1}$ denote the excitation energy and the angular momentum for the fragment $(Z_{1},N_{1})$, respectively, which are related to the center-of-mass energy $E_{c.m.}$ and incident angular momentum $J$. The maximal angular momentum $J_{max}$ is taken to be the grazing collision of two colliding nuclei. The capture cross section is given by $\sigma_{cap}=\pi\hbar^{2}(2J+1) T(E_{c.m.},J)/(2\mu E_{c.m.})$ and the probability $T(E_{c.m.},J)$ is calculated within the Hill-Wheeler formula.
For the heavy system, there is no potential pocket after overcoming the Coulomb barrier, e.g. the systems $^{136}$Xe+$^{208}$Pb, $^{238}$U+$^{198}$Pt etc. The classical trajectory approach is used with the relation of $T(E_{c.m.},J)=0$ and 1 for $E_{c.m.} < B+J(J+1)\hbar^{2}/(2\mu R^{2}_{C})$ and $E_{c.m.} > B+J(J+1)\hbar^{2}/(2\mu R^{2}_{C})$, respectively. The $\mu$ and $R_{C}$ denote the reduced mass and Coulomb radius by $\mu = m_{n} A_{p} A_{t} /(A_{p} + A_{t})$  with $m_{n}$, $A_{p}$ and $A_{t}$ being the nucleon mass and numbers of projectile and target nuclides, respectively. The distribution function is taken as the Gaussian form $f(B) = \frac{1}{N} exp[-((B-B_{m})/\Delta)^{2}]$, with the normalization constant satisfying the unity relation $\int f(B)dB=1$. The quantities $B_{m}$ and $\Delta$ are evaluated by $B_{m}=(B_{C} + B_{S})/2$ and $\Delta = (B_{C} - B_{S})/2$, respectively. The $B_{C}$ and $B_{S}$ are the Coulomb barrier at waist-to-waist orientation and the minimum barrier by varying the quadrupole deformation of the colliding partners.

The distribution probability is obtained by solving a set of master equations numerically in the potential energy surface of the DNS. The temporal evolution of the distribution probability $P(Z_{1},N_{1},E_{1},\beta_{1},B,t)$ for fragment 1 with proton number $Z_{1}$, neutron number $N_{1}$, excitation energy $E_{1}$, quadrupole deformation $\beta_{1}$ is described by the following master equations
\begin{eqnarray}
&&\frac{d P(Z_1,N_1,E_1,\beta_{1},B,t)}{d t} =  \sum \limits_{Z'_1=Z_1\pm1}  W_{Z_1,N_1,\beta_1; Z'_1, N_1, \beta^{\prime}_1}(t)  \nonumber \\
&& \times [d_{Z_1,N_1} P(Z'_1,N_1,E'_1,\beta^{\prime}_1,B,t) - d_{Z'_1,N_1}P(Z_1,N_1,E_1,\beta_1,B,t)]   \nonumber  \\
&& + \sum \limits_{N'_1=N_1\pm1} W_{Z_1,N_1,\beta_1; Z_1,N'_1,\beta^{\prime}_1}(t) [d_{Z_1,N_1} \nonumber   \\
&& P(Z_1,N'_1,E'_1,\beta^{\prime}_1,B,t) - d_{Z_1,N'_1}P(Z_1,N_1,E_1,\beta_1,B,t)]
\nonumber   \\
&& + \sum \limits_{Z'_1=\pm1,N'_1=N_1\pm1} W^{d}_{Z_1,N_1,\beta_1; Z'_1,N'_1,\beta^{\prime}_1}(t) [d_{Z_1,N_1}
\nonumber   \\
&&  P(Z'_1,N'_1,E'_1,\beta^{\prime}_1,B,t) - d_{Z'_1,N'_1}P(Z_1,N_1,E_1,\beta_1,B,t)]
\nonumber   \\
&& + \sum \limits_{Z'_1=\pm1,N'_1=N_1\pm2} W^{t}_{Z_1,N_1,\beta_1; Z'_1,N'_1,\beta^{\prime}_1}(t) [d_{Z_1,N_1}
\nonumber   \\
&& P(Z'_1,N'_1,E'_1,\beta^{\prime}_1,B,t) - d_{Z'_1,N'_1}P(Z_1,N_1,E_1,\beta_1,B,t)]
\nonumber   \\
&& + \sum \limits_{Z'_1=\pm2,N'_1=N_1\pm1} W^{^{3}He}_{Z_1,N_1,\beta_1; Z'_1,N'_1,\beta^{\prime}_1}(t) [d_{Z_1,N_1}
\nonumber   \\
&& P(Z'_1,N'_1,E'_1,\beta^{\prime}_1,B,t) - d_{Z'_1,N'_1}P(Z_1,N_1,E_1,\beta_1,B,t)]
\nonumber   \\
&& + \sum \limits_{Z'_1=\pm2,N'_1=N_1\pm2} W^{\alpha}_{Z_1,N_1,\beta_1; Z'_1,N'_1,\beta^{\prime}_1}(t) [d_{Z_1,N_1}
\nonumber   \\
&& P(Z'_1,N'_1,E'_1,\beta^{\prime}_1,B,t) - d_{Z'_1,N'_1}P(Z_1,N_1,E_1,\beta_1,B,t)].
\end{eqnarray}
Here the $W_{Z_{1},N_{1},\beta_{1}; Z^{\prime}_{1},N_{1},\beta^{\prime}_{1}}$ ($W_{Z_{1},N_{1},\beta_{1}; Z_{1},N^{\prime}_{1},\beta^{\prime}_{1}}$) is the mean transition probability from the channel ($Z_{1},N_{1}$) with the energy $E_{1}$ and quadrupole deformation $\beta_1$ to the DNS fragment ($Z^{\prime}_{1},N_{1}$) with $E^{\prime}_{1}$ and $\beta^{\prime}_1$ by transferring a proton, [or ($Z_{1},N_{1}$) to ($Z_{1},N^{\prime}_{1}$) by transferring a neutron]. The cluster transition probabilities by transferring deuteron, triton, $^{3}$He and $\alpha$ are denoted by $W^{d}_{Z_1,N_1,\beta_1; Z'_1,N'_1,\beta^{\prime}_1}(t)$, $W^{t}_{Z_1,N_1,\beta_1; Z'_1,N'_1,\beta^{\prime}_1}(t)$, $W^{^{3}He}_{Z_1,N_1,\beta_1; Z'_1,N'_1,\beta^{\prime}_1}(t)$ and $W^{\alpha}_{Z_1,N_1,\beta_1; Z'_1,N'_1,\beta^{\prime}_1}(t)$ from the channel ($Z_{1},N_{1}$) to ($Z'_{1},N'_{1}$), respectively.

The microscopic dimension $d_{Z_{1},N_{1}}$ corresponds to the DNS fragment ($Z_{1},N_{1},E_{1},\beta_1$). The nucleon or cluster transfer is considered by the relations of $Z^{'}_{1}=Z_{1}\pm Z_{\nu}$ and $N^{'}_{1}=N_{1}\pm N_{\nu}$ with the proton number $Z_{\nu}$ and neutron number $N_{\nu}$, in which the nucleon transfer by distinguishing effect, deuteron, triton, $^{3}$He and $\alpha$ are taken into account in the process of solving the master equations. At the initial time, the distribution probabilities of projectile and target nuclei are set to be $P(Z_{proj},N_{proj},E_{1}=0,\beta_{1}=\beta_{proj},B,t=0)=$0.5 and $P(Z_{targ},N_{targ},E_{1}=0,\beta_{1}=\beta_{targ},B,t=0)=$0.5. The unitary condition is also satisfied by the relation of $\sum_{Z_{1},N_{1}} P(Z_{1}, N_{1}, E_{1},\beta_{1},B,t)=1$ during the time evolution in the relaxation process after the inclusion of cluster transfer.

The transfer of a nucleon or cluster in the interacting nuclei is described by the single-particle Hamiltonian
\begin{equation}
H(t)=H_{0}(t)+V(t)
\end{equation}
with
\begin{eqnarray}
 H_{0}(t) && = \sum_{K}\sum_{\nu_{K}}\varepsilon_{\nu_{K}}(t)a_{\nu_{K}}^{\dag}(t)a_{\nu_{K}}(t),   \\
 V(t) && = \sum_{K,K^{\prime}}\sum_{\alpha_{K},\beta_{K^{\prime}}}u_{\alpha_{K},\beta_{K^{\prime}}}(t)
a_{\alpha_{K}}^{\dag}(t)a_{\beta_{K^{\prime}}}(t)        \nonumber \\
&& =\sum_{K,K^{\prime}}V_{K,K^{\prime}}(t).
\end{eqnarray}
Here the indices $K, K^{\prime}$ $(K,K^{\prime}=1,2)$ denote the DNS fragments $1$ and $2$. The quantities $\varepsilon_{\nu_{K}}$ and $u_{\alpha_{K},\beta_{K^{\prime}}}$ are the single-particle energies and the interaction matrix elements, respectively. The single particle states are defined with respect to the centers of
the interacting nuclei and are assumed to be orthogonalized in the overlap region. So the annihilation and creation operators are dependent on the local excitation energy. The single particle matrix elements are parameterized by
\begin{eqnarray}
u_{\alpha_{K},\beta_{K^{\prime}}}(t) && = U_{K,K^{\prime}}(t) \exp\left[-\frac{1}{2}
\left(\frac{\varepsilon_{\alpha_{K}}(t)-\varepsilon_{\beta_{K^{\prime}}}(t)}
{\Delta_{K,K^{\prime}}(t)}\right)^{2}\right]         \nonumber \\
&& - U_{K,K^{\prime}}(t)\delta_{\alpha_{K},\beta_{K^{\prime}}}
\end{eqnarray}
with the energy width $\Delta_{K,K^{\prime}}(t)=2$ MeV. The strength parameters are given by
\begin{equation}
U_{K,K^{\prime}}(t) = \frac{g_{1}^{1/3}g_{2}^{1/3}}{g_{1}^{1/3}+g_{2}^{1/3}} \frac{2\gamma_{KK^{\prime}}}{g_{K}^{1/3}g_{K^{\prime}}^{1/3}}
\end{equation}
with the values of $\gamma_{11}=\gamma_{22}=\gamma_{12}=\gamma_{21}=3$.

The excited DNS opens a valence space in which the valence nucleons have a symmetrical distribution around the Fermi surface. Only the particles at the states within the valence space are actively at excitation and transfer. The averages on these quantities are performed in the valence space as follows.
\begin{eqnarray}
\Delta \varepsilon_K = \sqrt{\frac{4\varepsilon^*_K}{g_K}},\quad
\varepsilon^*_K =\varepsilon^*\frac{A_K}{A}, \quad
g_K = A_K /12,
\end{eqnarray}
where the $\varepsilon^*$ is the local excitation energy of the DNS. The microscopic dimension for the fragment ($Z_{K},N_{K}$) is evaluated by the valence states $N_K$ = $g_K\Delta\varepsilon_K$ and the valence nucleons $m_K$ = $N_K/2$ ($K=1,2$) as
\begin{eqnarray}
 d(m_1, m_2) = {N_1 \choose m_1} {N_2 \choose m_2}.
\end{eqnarray}

The transition probabilities of neutron, proton, deuteron, triton, $^{3}$He and $\alpha$ are related to the local excitation energy, which are microscopically estimated from the interaction potential in valence space as
\begin{eqnarray}
W^{\nu}_{Z_{1},N_{1}; Z_{1}^{\prime},N_{1}^{\prime}}=&& G_{\nu} \frac{\tau_{mem}(Z_{1},N_{1},E_{1}; Z_{1}^{\prime},N_{1}^{\prime},E_{1}^{\prime})} {d_{Z_{1},N_{1}} d_{Z_{1}^{\prime},N_{1}^{\prime}}\hbar^{2}}  \times   \nonumber \\
&& \sum_{ii^{\prime}}| \langle  i^{\prime}| V | i \rangle |^{2}
\end{eqnarray}
with $Z_{1}^{\prime}=Z_{1}\pm Z_{\nu}$ and $N_{1}^{\prime}=N_{1}\pm N_{\nu}$. The $Z_{\nu}$ and $N_{\nu}$ are the proton and neutron numbers of transferring a cluster, respectively. The spin-isospin statistical factors $G_{\nu}$ are taken to be 1, 1, 3/8, 1/12, 1/12 and 1/96 corresponding to proton, neutron, deuteron, triton, helium-3 and alpha, respectively, which relies on the Wigner density approach for recognizing the clusters in heavy-ion collisions \cite{Ma97,Fe20}.

The memory time is related to the interaction potential and estimated by \cite{Wo78}
\begin{equation}
\tau_{mem}(Z_{1},N_{1},E_{1}; Z_{1}^{\prime},N_{1}^{\prime},E_{1}^{\prime}) = \left[\frac{2\pi \hbar^2} {\sum _{KK'} <V_{KK} V^*_{KK'}>}\right]^{1/2},
\end{equation}
\begin{eqnarray}
<V_{KK} V^*_{KK'}> && =  \frac{1}{4} U^2_{KK'}g_{K}g_{K^{\prime}} \Delta_{KK'} \Delta \varepsilon_K \Delta \varepsilon_{K^{\prime}}            \nonumber \\
&& \times \left[ \Delta^2_{KK'}+ \frac{1}{6} ((\Delta \varepsilon_K)^2  + (\Delta \varepsilon_{K^{\prime}})^2) \right] ^{-1/2}
\end{eqnarray}
The interaction matrix element with transferring a cluster is given by
\begin{eqnarray}
\sum_{ii^{\prime}}| \langle  i^{\prime}| V | i \rangle |^{2}  =  \omega_{11}(ii) + \omega_{22}(i^{\prime}i^{\prime})  + \omega_{12}(ii^{\prime}) + \omega_{21}(i^{\prime}i),
\end{eqnarray}
in which the element is calculated by
\begin{equation}
\omega_{KK'} (i, i^{\prime})=d_{Z_1,N_1} <V_{KK'} V^*_{KK'}>
\end{equation}
with the states $i(Z_{1},N_{1},E_{1})$ and $i^{\prime}(Z_{1}^{\prime},N_{1}^{\prime},E_{1}^{\prime})$.
In the relaxation process of the relative motion, the DNS will be excited by the dissipation of the relative kinetic energy. The local excitation energy is determined by the dissipation energy from the relative motion and the potential energy surface of the DNS \cite{Fe07}.

\begin{figure*}
\includegraphics[width=16 cm]{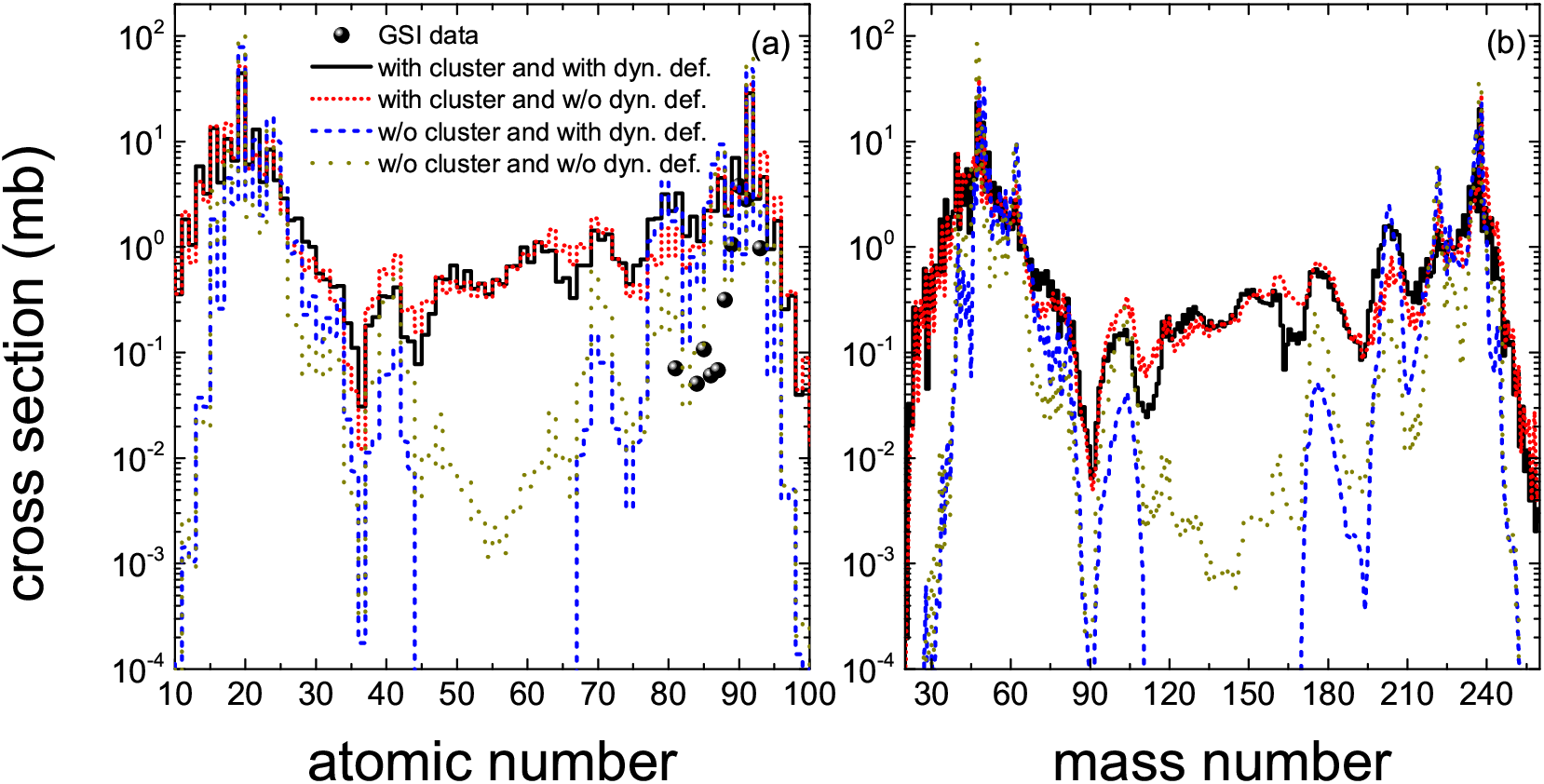}
\caption{ Comparison of the MNT fragments in collisions of $^{48}$Ca + $^{238}$U at the beam energy of 4.76 MeV/nucleon with different cases of cluster transfer (deuteron, triton, $^{3}$He and $\alpha$) and dynamical evolution of quadrupole deformation of DNS fragments. The available experimental data are taken from \cite{De20}. }
\end{figure*}

The MNT reactions have been extensively investigated both in experiments and in theories. The reaction mechanism has been considered as a unique way to reach the neutron-rich heavy nuclei even the 'island of stability' of superheavy nuclides in nuclear chart, i.e., around the neutron shell closure with N=126, 152 and 162. The properties of neutron-rich heavy nuclei have significant implications in understanding the nucleosynthesis in universe beyond the iron element, shell evolution of isotopes and isotones, new fission mechanism, new superheavy nuclides or elements, nuclear spectroscopies etc. To describe the MNT reactions in theories, sophisticated transport models are still expected for the collision dynamics. The dinuclear system model has been extensively used for estimating the SHN production cross section in the cold fusion reactions with the $^{208}$Pb and $^{209}$Bi targets, $^{48}$Ca induced fusion reactions and for modeling the MNT reaction dynamics. As a test of the DNS model, the available data at GSI in collisions of $^{48}$U+$^{248}$Cm at the beam energy of E$_{lab}$ = 4.76 MeV/nucleon (E$_{c.m.}$=190.1 MeV, $V_{tip-tip}$=172.1 MeV and $V_{waist-waist}$=184.6 MeV) are compared with the calculations as shown in Fig. 1. It should be noticed that the products of MNT fragments were populated by the differential cross sections at the SHIP acceptance angle of (0$\pm$2) degrees towards to the beam direction \cite{De20}. The effects of cluster transfer and dynamical quadrupole deformation are coupled to the dissipation of relative motion energy and angular momentum. The bump structure of the charged numbers of MNT fragments is caused from the shell effects, i.e., with the numbers of Z=28, 50 and 82, which is favorable for the fragment formation because of the larger distribution probability and survival against the particle evaporation. Overall, the broader charge and mass distributions are obtained with the cluster transfer taken into account in the DNS model. It should be noticed that the consideration of sole nucleon transfer and dynamical deformation enable the nearly symmetric distributions of the charge and mass spectra. The cluster transfer is more pronounced in the target-like region (heavier fragments) and leads to the broader mass distribution. The overestimation of the production cross sections in the target-like region in comparison with the GSI data is caused from the forward measurements of MNT fragments. It is well known that the MNT fragments manifest the anisotropic distribution and the angular distributions of the projectile-like and target-like fragments are related to the beam energy \cite{Ko12,Ch17,Pe22}.

\begin{figure*}
\includegraphics[width=16 cm]{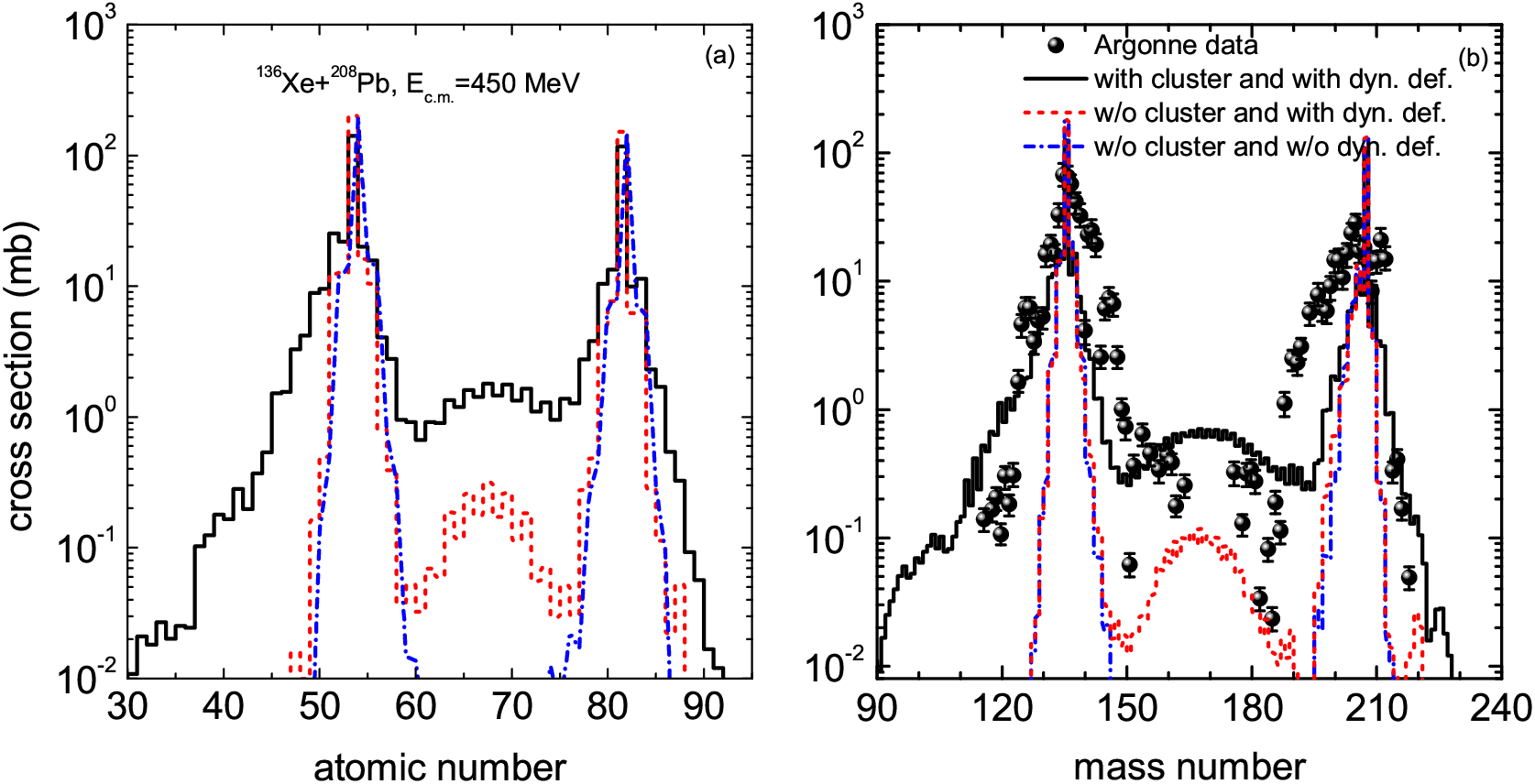}
\includegraphics[width=16 cm]{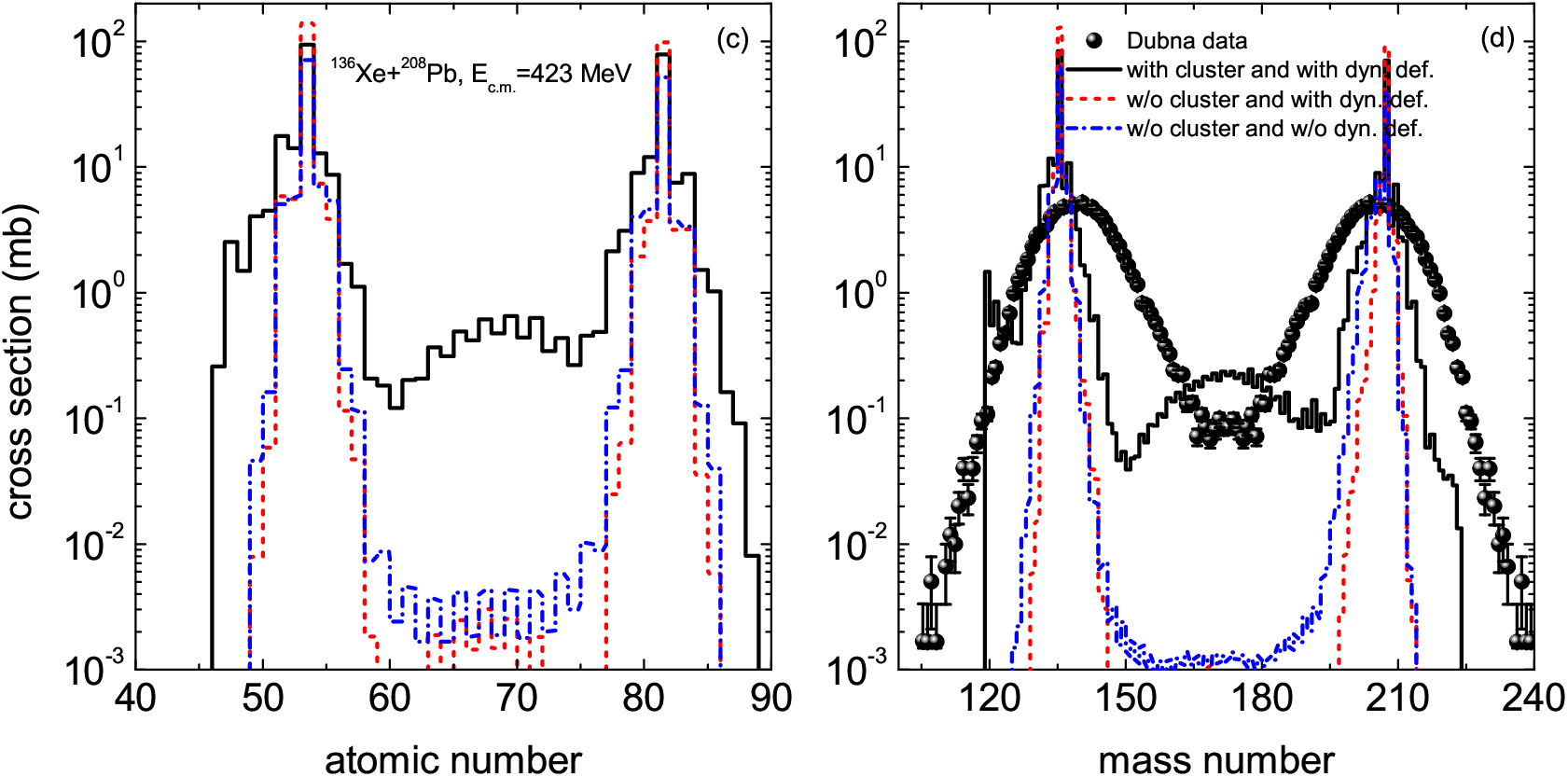}
\caption{Production cross sections of fragments as functions of atomic and mass numbers in the $^{136}$Xe+$^{208}$Pb reaction at the center of mass energies of E$_{c.m.}$ =450 MeV and 423 MeV, respectively. The available experimental data are measured at Dubna \cite{Ko12} and Argonne \cite{Ba15}. }
\end{figure*}

The neutron closed shell N=126 is particular significant for stabilizing and elongating the lifetime of neutron-rich isotopes via the MNT reactions, which plays an essential role for the production of heavy elements beyond the iron element in stellar nucleosynthesis through the \emph{r}-process at the 'waiting point'. The low-energy MNT reaction of $^{136}$Xe+$^{208}$Pb for production of new heavy isotopes was proposed by Zagrebaev and Greiner for the first time with the multidimensional Langevin approach \cite{Za08}. Accurate estimation of production cross sections in the MNT reactions is still expected, also the systems of $^{136}$Xe + $^{198}$Pt/$^{197}$Au/$^{204}$Hg/$^{205}$Tl. The influence of cluster transfer and dynamical deformation on the MNT fragment formation in the reaction of $^{136}$Xe+$^{208}$Pb is analyzed as shown in Fig. 2. The incident energies of 423 MeV at Dubna \cite{Ko12} and 450 MeV at Argonne \cite{Ba15} are selected for comparison corresponding to below and above the Coulomb barrier ($V_{C}$=427.3 MeV), respectively. The inclusion of dynamical deformation is obvious in the region of symmetric mass fragments at the center of mass energy of 450 MeV. The cluster transfer broadens the charge and mass distributions. It should be noticed that the total cross sections of all MNT fragments for the three cases are equal to the capture cross sections, e.g., 82 mb and 69 mb for $^{136}$Xe and $^{208}$Pb with the cluster transfer and dynamical deformation, respectively, 129 mb and 90 mb for $^{136}$Xe and $^{208}$Pb solely with the dynamical deformation, 62 mb and 38 mb for $^{136}$Xe and $^{208}$Pb without inclusion of the cluster transfer and dynamical deformation at the energy of 423 MeV in the calculation. The primary fragments manifest the symmetric distributions with the atomic number or mass number.
The results with the cluster transfer and dynamical deformation are consistent with the Argonne data. But at the energy below the Coulomb barrier, the narrower mass distributions are obtained in comparison with the Dubna data, which might be caused from the fact the shape evolution is neglected and individual properties are assumed in the DNS model. The shape evolution becomes important in the low-energy nuclear collisions, which is associated with the neck formation and disappearance of entrance channel information with the configuration evolution of colliding system.

\begin{figure*}
\includegraphics[width=16 cm]{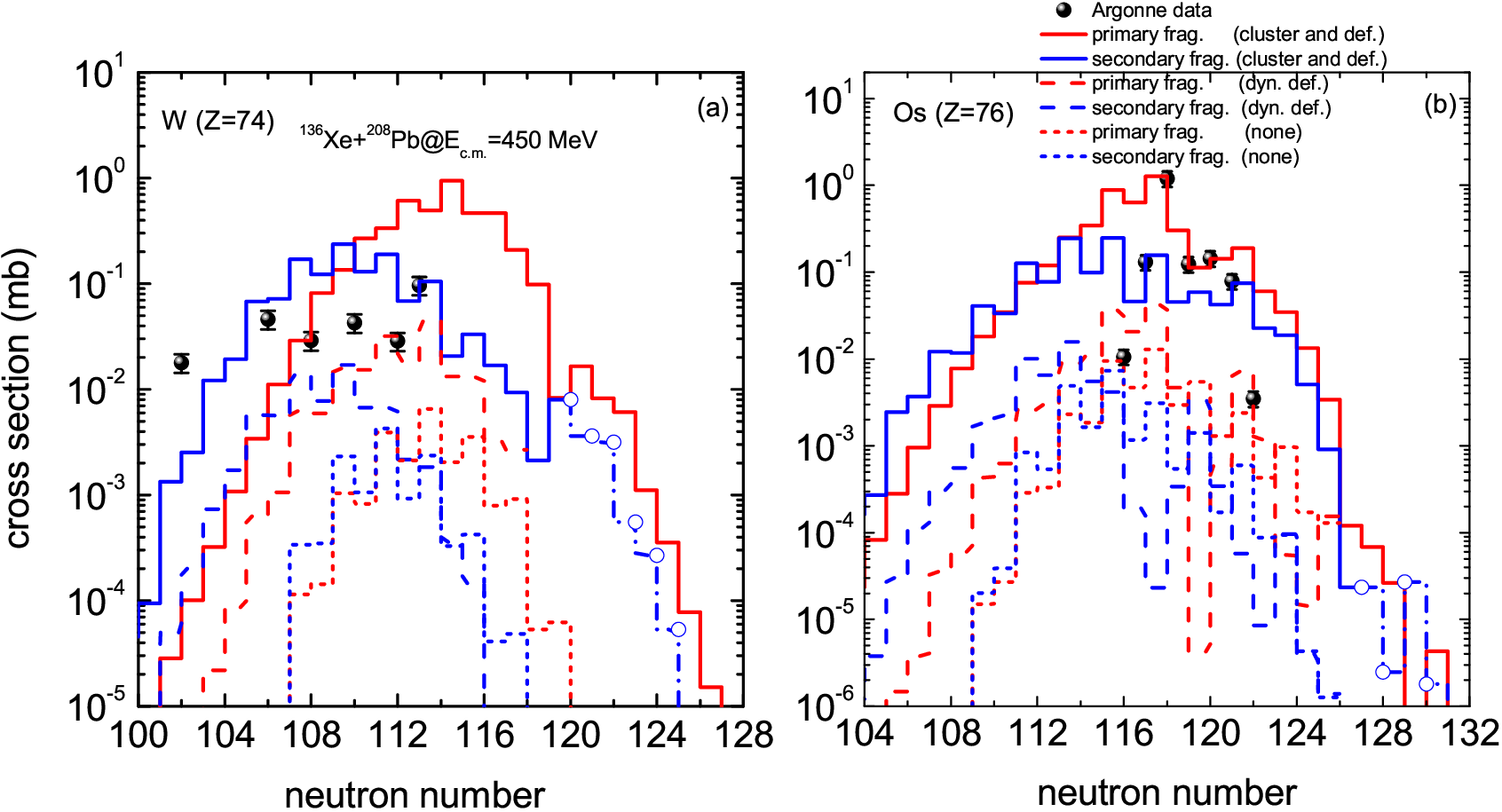}
\caption{Isotopic distributions of the primary and secondary fragments for wolfram (W) and osmium (Os) production in the MNT reactions of $^{136}$Xe+$^{208}$Pb at E$_{c.m.}$ = 450 MeV with the effects of cluster transfer and dynamical deformation. The open symbols denote the new isotopes \cite{Wa21}. }
\end{figure*}

The advantage of MNT reactions is to create the neutron-rich heavy isotopes, in particular around the neutron closed shell. The isotopic and isotonic distributions of MNT products are of importance to investigate the shell evolution with varying the neutron and proton numbers of fragments, nuclear spectroscopies via the decay modes, reaction dynamics associated with the neck evolution, nucleon or cluster transfer, temporal evolutions of deformation parameters (quadrupole, octupole and hexadecapole). Figure 3 shows the isotopic distributions of the primary (red lines) and secondary (blue lines) fragments for wolfram (W) and osmium (Os) production in the MNT reactions of $^{136}$Xe+$^{208}$Pb at E$_{c.m.}$ = 450 MeV. It is obvious that the primary fragments manifest the larger cross sections of the neutron-rich isotope production, e.g., the maximal position around the isotopes $^{188}$W and $^{194}$Os for the 8 and 6 proton pickup reactions. The secondary decay by evaporating several neutrons enables the maximal yields close to the $\beta-$stability line. The inclusion of the cluster transfer and dynamical deformation leads to the broader isotope distribution with the cross sections above 10 nb and is more consistent with the Argonne data \cite{Ba15}. The new isotopes might be created via the MNT reactions with the possible measurements in experiments \cite{Wa21}, i.e., from 8$\mu$b for $^{194}$W to 4.5 nb for $^{200}$W, and from 23.5 nb for $^{203}$Os to 1.8 nb for $^{206}$Os. The stripping reactions are also investigated for the isotopic distribution of radon (Rn) and francium (Fr) production in the MNT reactions of $^{136}$Xe + $^{208}$Pb as shown in Fig. 4. The 4 and 5 stripping protons enable the checking on the neutron shell evolution of N=126. The available data of radon isotope production are nicely reproduced via the secondary decay spectrum (blue line) by including the cluster transfer and dynamical deformation in the DNS model. A narrower isotopic distribution of francium production can not be reproduced by the model, i.e., 15.9 $\mu$b for $^{220}$Rn and 29.7 $\mu$b for $^{220}$Fr in the calculations, but the lower cross section of 5.7$\pm$1.1 $\mu$b for $^{216}$Fr in experiments.

\begin{figure*}
	\includegraphics[width=16 cm]{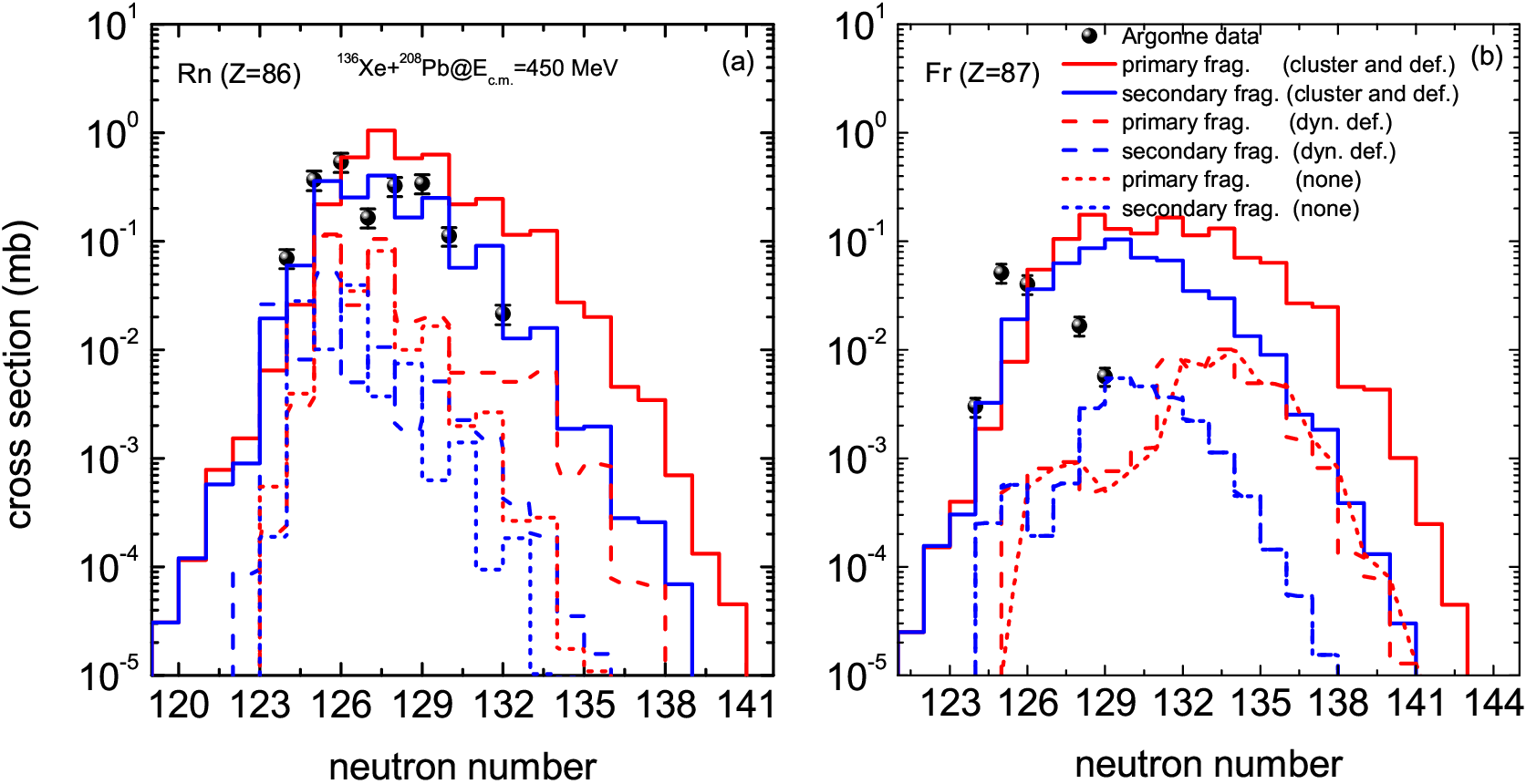}
	\caption{ Production cross sections of radon (Rn) and francium (Fr) production in the MNT reactions of $^{136}$Xe + $^{208}$Pb at E$_{c.m.}$ = 450 MeV and compared with the Argonne data \cite{Ba15}. }
\end{figure*}

The closed neutron shell of N=126 and proton shell of Z=82 are favorable for the MNT fragment formation. The nuclear spectroscopies of shell evolution, in particular for the nuclide properties beyond the $\beta$-stability line, are of importance for exploring the nucleosynthesis in the \emph{r}-process, i.e., the heavy element creation in the binary neutron star merging. In the terrestrial laboratories, the neutron-rich heavy isotopes might be created via the MNT reactions. The shell effect will enhance the fission barrier and enlarges the separation energy of the MNT fragments, which are favorable for the primary products and for the survival of the cold fragments via the binary fission and $\beta$-decay. The isotonic and isotopic distributions of the MNT products formed in the reactions of $^{58,64,72}$Ni + $^{198}$Pt near the Coulomb barrier energies as shown in Fig. 5. It is obvious that the system $^{72}$Ni + $^{198}$Pt ($V_{C}$=212.2 MeV) is available for the neutron-rich isotope production, e.g., the maximal cross section with 0.23 mb for $^{208}$Pb, 2.6 nb for $^{202}$Os. The reaction of $^{58}$Ni + $^{198}$Pt ($V_{C}$=228.2 MeV) is favorable for the proton-rich isotope, e.g., 13.1 nb for $^{196}$Pb and 0.12 nb for $^{193}$Pb. The incident energy dependence of $^{64}$Ni + $^{198}$Pt ($V_{C}$=220.3 MeV) is obvious for the proton-rich isotope production.

\begin{figure*}
	\includegraphics[width=16 cm]{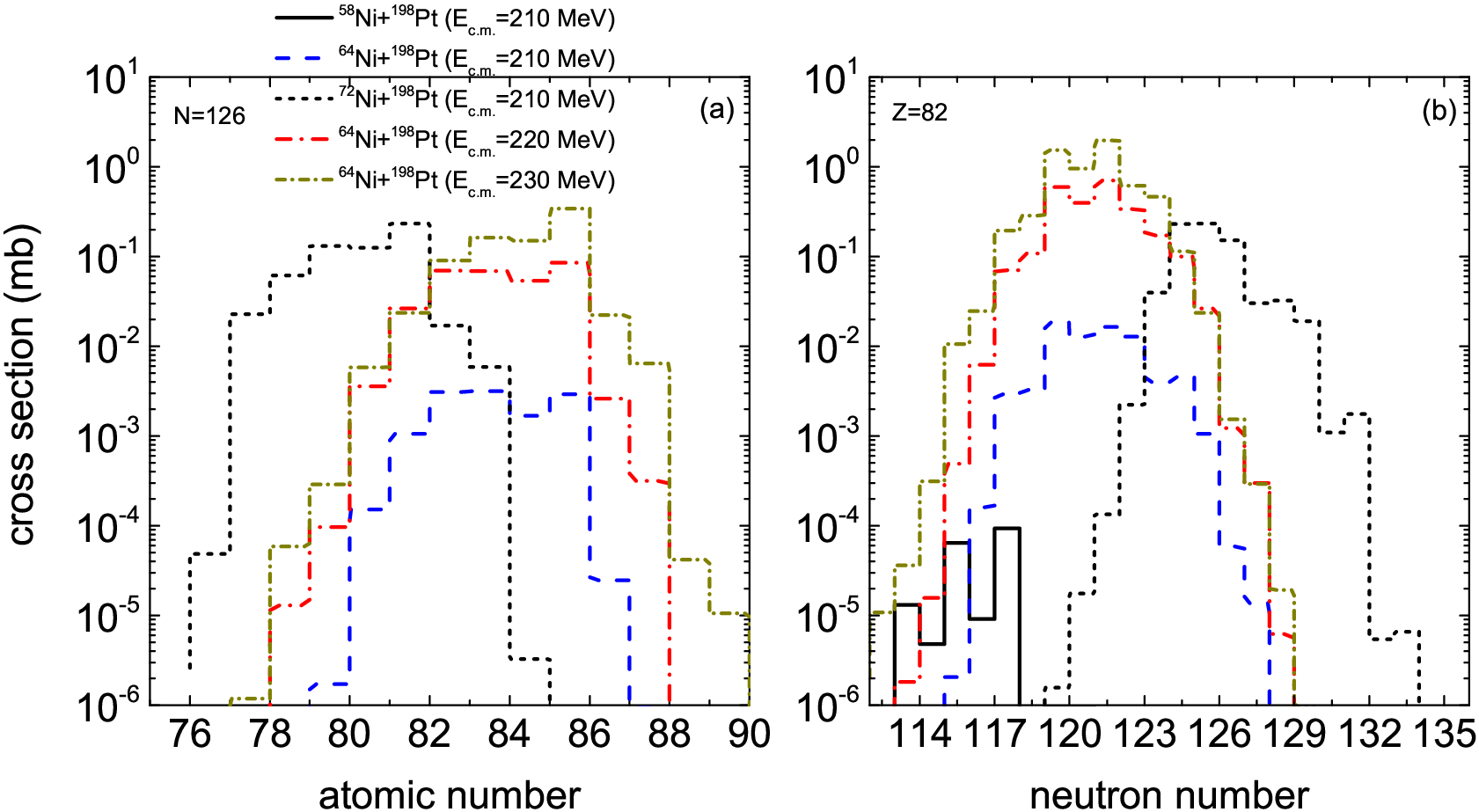}
	\caption{ The secondary fragment production in the MNT reactions of $^{58,64,72}$Ni + $^{198}$Pt around the shell closure with N=126 (left panel) and Z=82 (right panel), respectively. }
\end{figure*}

The beam energy influences the dissipation energy into the DNS and consequently the MNT fragment formation. The more local excitation energy is obtained with increasing the incident energy and leads to the wider domain nucleon transfer \cite{Ni17}. A number of rare isotopes might be created with more energetic nuclear collisions. However, the less survival of the primary fragment is obtained by the transfer dynamics. The competition of the diffusion of nucleon transfer and the survival of formed fragment leads the isotopic structure of final MNT fragments, which is associated with the beam energy and colliding system. It is found that the total mass and charge distributions of the secondary fragments around the shell closure \emph{N }= 126 weakly depend on the bombarding energy in the MNT reaction of $^{136}$Xe + $^{198}$Pt \cite{Fe17}. Systematic investigation of energy dependence on the fragment formation in the MNT reactions would be helpful for selecting the optimal beam energy in experiment. Shown in Fig. 6 is the isotopic distributions of platinum, iridium, osmium and rhenium  in the pick-up reactions of $^{64}$Ni + $^{198}$Pt and $^{72}$Ni + $^{198}$Pt at the center of mass energies of 210 MeV, 220 MeV and 230 MeV, respectively. Overall, the neutron-rich radioactive nuclide $^{72}$Ni induced reactions are favorable for the neutron-rich isotope production, in particular at the incident energies 220 MeV and 230 MeV above the Coulomb barrier ($V_{C}$=212.2 MeV). The maximal yields of the isotopic distributions deviate from the $\beta$-stability line to the neutron-rich region, i.e., 27.8 mb, 1.58 mb, 3.03 mb and 0.82 mb for $^{198}$Pt, $^{197}$Ir, $^{194}$Os and $^{191}$Re, respectively.

\begin{figure*}
	\includegraphics[width=16 cm]{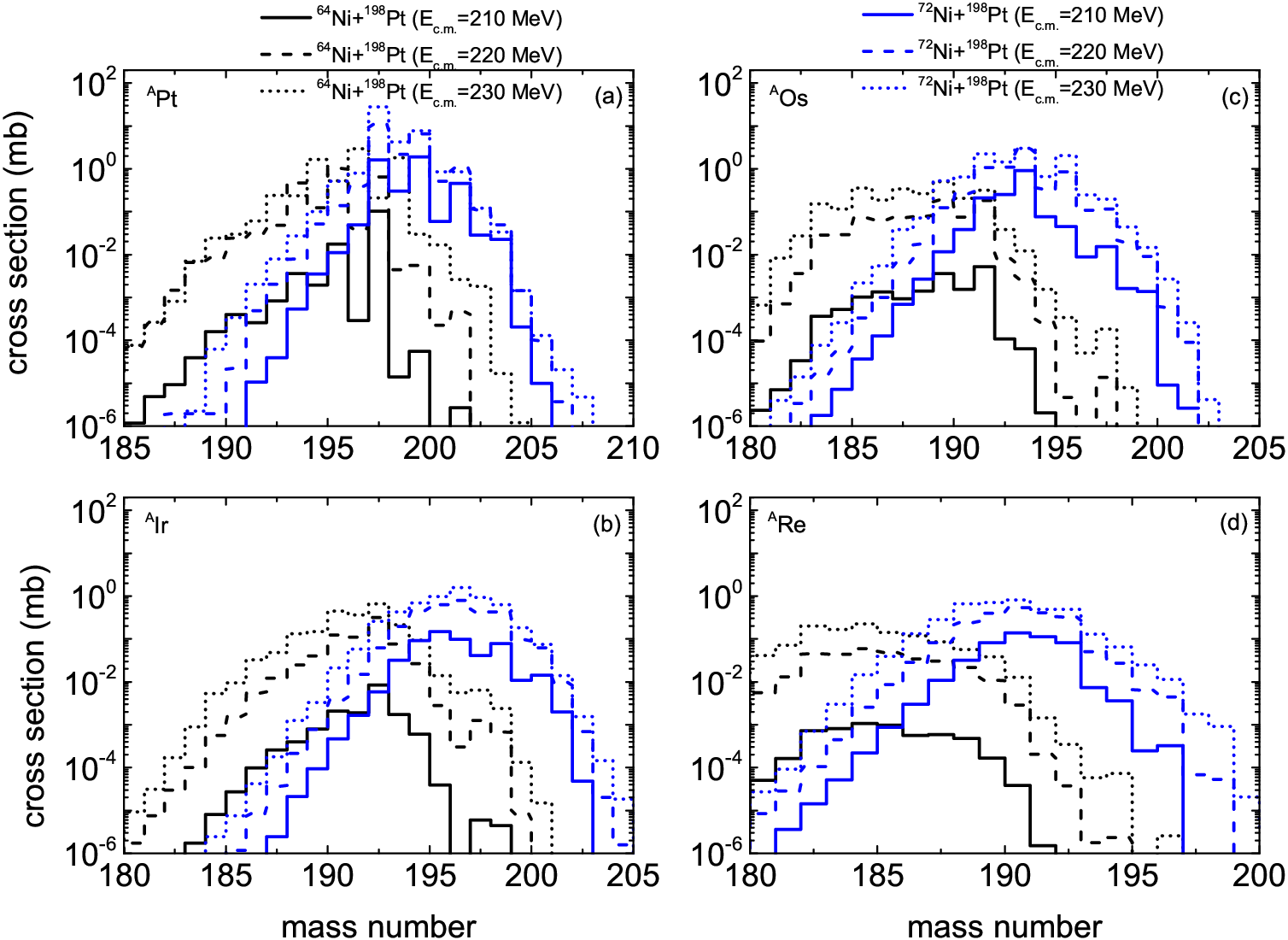}
	\caption{ Isotopic distributions of platinum (Pt), iridium (Ir), osmium (Os) and rhenium (Re) in the MNT reactions of $^{64}$Ni + $^{198}$Pt and $^{72}$Ni + $^{198}$Pt at the center of mass energies of 210 MeV, 220 MeV and 230 MeV, respectively. }
\end{figure*}

In summary, the DNS model is improved by implementing the cluster transfer into the master equations, i.e., deuteron, triton, $^{3}$He and $\alpha$, in which the nucleon transfer and cluster effect are coupled to the temporal evolution of quadrupole deformation and dissipation of the relative motion energy and angular momentum. The inclusion of cluster transfer in the DNS model is favorable for the MNT fragment formation and leads to a broad isotopic distribution and the distribution probability is associated with the cluster separation energy of DNS fragment. The production cross sections of the MNT fragments over transferring 20 nucleons are nicely consistent with the Argonne data. The neutron-rich isotopes of elements W and Os around N=126 are predicted with the cross sections above 10 nb in the MNT reactions of $^{136}$Xe+$^{208}$Pb at E$_{c.m.}$ = 450 MeV. The system $^{72}$Ni + $^{198}$Pt is favorable for the production of neutron-rich nuclides of isotones with N=126 and of isotopes with Z=82. The production cross sections of isotones and proton-rich isotopes are associated with the beam energy. However, the neutron-rich nuclides are nearly independent on the beam energy.

\textbf{Acknowledgements}
This work was supported by the National Natural Science Foundation of China (Projects No. 12175072 and No. 11722546) and the Talent Program of South China University of Technology (Projects No. 20210115).

\end{document}